\documentclass[11pt]{article}
\usepackage{epsfig,fancyheadings}\usepackage{redner}

\makeatletter
\renewcommand\theequation{\thesection.\arabic{equation}}
\@addtoreset{equation}{section}
\makeatother

\begin{document}
\title{A Statistical Physics Perspective on Web Growth}
\author{P.~L.~Krapivsky and S.~Redner\\
Center for BioDynamics, Center for Polymer Studies,\\
and Department of Physics, Boston University, Boston, MA 02215, USA}
\maketitle
\begin{abstract}  
  Approaches from statistical physics are applied to investigate the
  structure of network models whose growth rules mimic aspects of the
  evolution of the world-wide web.  We first determine the degree
  distribution of a growing network in which nodes are introduced one at a
  time and attach to an earlier node of degree $k$ with rate $A_k\sim
  k^\gamma$.  Very different behaviors arise for $\gamma<1$, $\gamma=1$, and
  $\gamma>1$.  We also analyze the degree distribution of a heterogeneous
  network, the joint age-degree distribution, the correlation between degrees
  of neighboring nodes, as well as global network properties.  An extension
  to directed networks is then presented.  By tuning model parameters to
  reasonable values, we obtain distinct power-law forms for the in-degree and
  out-degree distributions with exponents that are in good agreement with
  current data for the web.  Finally, a general growth process with
  independent introduction of nodes and links is investigated.  This leads to
  independently growing sub-networks that may coalesce with other
  sub-networks.  General results for both the size distribution of
  sub-networks and the degree distribution are obtained.
  
\end{abstract}
  
\section{Introduction}

With the recent appearance of the Internet and the world-wide web,
understanding the properties of growing networks with popularity-based
construction rules has become an active and fruitful research area
\cite{review}.  In such models, newly-introduced nodes preferentially attach
to pre-existing nodes of the network that are already ``popular''.  This
leads to graphs whose structure is quite different from the well-known {\em
  random graph} \cite{bol,jan} in which links are created at random between
nodes without regard to their popularity.  This discovery of a new class of
graph theory problems has fueled much effort to characterize their
properties.

One basic measure of the structure of such networks is the {\em node degree}
$N_k$ defined as the number of nodes in the network that are linked to $k$
other nodes.  In the case of the random graph, the node degree is simply a
Poisson distribution.  In contrast, many popularity-driven growing networks
have much broader degree distributions with a stretched exponential or a
power-law tail.  The latter form means that there is no characteristic scale
for the node degree, a feature that typifies many networked systems
\cite{review}.

Power laws, or more generally, distributions with highly skewed tails,
characterize the degree distributions of many man-made and naturally
occurring networks \cite{review}.  For example, the degree distributions at
the level of autonomous systems and at the router level exhibit highly skewed
tails \cite{fff,matta,as}.  Other important Internet-based graphs, such as
the hyperlink graph of the world-wide web also appear to have a degree
distribution with a power-law tail \cite{kum,BA,www1,www2,www3}.  These
observations have spurred a flurry of recent work to understand the
underlying mechanisms for these phenomena.

A related example with interest to anyone who publishes, is the distribution
of scientific citations \cite{lotka,LS,redner}.  Here one treats publications
as nodes and citations as links in a citation graph.  Currently-available
data suggests that the citation distribution has a power-law tail with an
associated exponent close to $-3$ \cite{redner}.  As we shall see, this
exponent emerges naturally in the {\it Growing Network} (GN) model where the
relative probability of linking from a new node to a previous node
(equivalent to citing an earlier paper) is strictly proportional to the
popularity of the target node.

In this paper, we apply tools from statistical physics, especially the rate
equation approach, to quantify the structure of growing networks and to
elucidate the types of geometrical features that arise in networks with
physically-motivated growth rules.  The utility of the rate equations has
been demonstrated in a diverse range of phenomena in non-equilibrium
statistical physics, such as aggregation \cite{agg}, coarsening
\cite{coarse}, and epitaxial surface growth \cite{surf}.  We will attempt to
convince the reader that the rate equations are also a simple yet powerful
analysis tool to analyze growing network systems.  In addition to providing
comprehensive information about the node degree distribution, the rate
equations can be easily adapted to analyze both heterogeneous and directed
networks, the age distribution of nodes, correlations between node degrees,
various global network properties, as well as the cluster size distribution
in models that give rise to independently evolving sub-networks.  Thus the
rate equation method appears to be better suited for probing the structure of
growing networks compared to the classical approaches for analyzing random
graphs, such as probabilistic \cite{bol} or generating function \cite{jan}
techniques.

In the next section, we introduce three basic models that will be the focus
of this review.  In the following three sections, we then present rate
equation analyses to determine basic geometrical properties of these
networks.  We close with a brief summary.

\section{Models}

The models we study appear to embody many of the basic growth processes in
web graphs and related systems.  These include:

\begin{itemize}
  
\item The {\em Growing Network} (GN) \cite{BA,simon}.  Nodes are added one at
  a time and a single link is established between the new node and a
  pre-existing node according to an attachment probability $A_k$ that depends
  only on the degree of the ``target'' node (Fig.~\ref{network}).

\begin{figure}[ht]
  \begin{center}
    \includegraphics[width=0.3\textwidth]{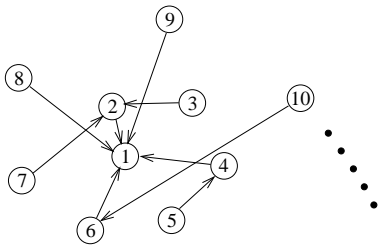}
 \caption{Growing network.  Nodes are added sequentially and 
   a single link joins a new node to an earlier node.  Node 1 has (total)
   degree 5, node 2 has degree 3, nodes 4 and 6 have degree 2, and the
   remaining nodes have degree 1.}~\label{network}
  \end{center}
\end{figure}

\item The {\em Web Graph} (WG).  This represents an extension of the GN to
  incorporate link directionality \cite{KRR} and leads to independent,
  dynamically generated in-degree and out-degree distributions.  The network
  growth occurs by two distinct processes \cite{gen} that are meant to mimic
  how hyperlinks are created in the web (Fig.~\ref{io-growth}):

\begin{itemize}
\item[(i)] With probability $p$, a new node is introduced and it immediately
  attaches to an earlier target node.  The attachment probability depends
  only on the in-degree of the target.
\item[(ii)] With probability $q=1-p$, a new link is created between already
  existing nodes.  The choices of the originating and target nodes depend on
  the out-degree of the former and the in-degree of the latter.
\end{itemize}

\begin{figure}[ht]
  \begin{center}
    \includegraphics[width=0.35\textwidth]{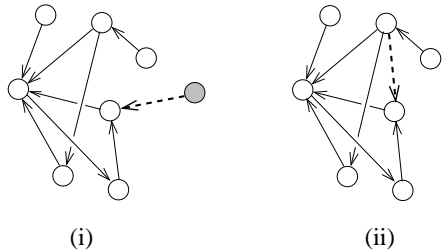}
\caption{Growth processes in the web graph model: 
  (i) node creation and immediate attachment, and (ii) link creation.  In (i)
  the new node is shaded, while in both (i) and (ii) the new link is dashed.}
\label{io-growth}
\end{center}
\end{figure}

\item The {\em Multicomponent Graph} (MG).  Nodes and links are introduced
  {\em independently} \cite{clusters}.  (i) With probability $p$, a new {\em
    unlinked} node is introduced, while (ii) with probability $q=1-p$, a new
  link is created between existing nodes.  As in the WG, the choices of the
  originating and target nodes depend on the out-degree of the former and the
  in-degree of the latter.  Step (i) allows for the formation of many
  clusters.

\end{itemize}

\section{Structure of the Growing Network}

Because of its simplicity, we first study the structure of the GN
\cite{BA,simon}.  The basic approaches developed in this section will then be
extended to the WG and MG models.

\subsection{Degree Distribution of a Homogeneous Network}

We first focus on the node degree distribution $N_k$.  To determine its
evolution, we shall write the rate equations that account for the change in
the degree distribution after each node addition event.  These equations
contain complete information about the node degree, from which any measure of
node degree (such as moments) can be easily extracted.  For the GN growth
process in which nodes are introduced one at a time, the rate equations for
the degree distribution $N_k(t)$ are \cite{KRL}
\begin{equation}
\label{Nk}
{d N_k\over dt}=
{A_{k-1} N_{k-1}-A_k N_k\over A}+\delta_{k1}.
\end{equation}
The first term on the right, $A_{k-1}N_{k-1}/A$, accounts for processes in
which a node with $k-1$ links is connected to the new node, thus increasing
$N_k$ by one.  Since there are $N_{k-1}$ nodes of degree $k-1$, the rate at
which such processes occur is proportional to $A_{k-1}N_{k-1}$, and the
factor $A(t)=\sum_{j\geq 1} A_jN_j(t)$ converts this rate into a normalized
probability.  A corresponding role is played by the second (loss) term on the
right-hand side; $A_kN_k/A$ is the probability that a node with $k$ links is
connected to the new node, thus leading to a loss in $N_k$.  The last term
accounts for the introduction of new nodes with no incoming links.

We start by solving for the time dependence of the moments of the degree
distribution defined via $M_n(t)=\sum_{j\geq 1} j^n N_j(t)$.  This is a
standard method of analysis of rate equations by which one can gain partial,
but valuable, information about the time dependence of the system with
minimal effort.  By explicitly summing Eqs.~(\ref{Nk}) over all $k$, we
easily obtain $\dot M_0(t)=1$, whose solution is $M_0(t)= M_0(0)+t$.  Notice
that by definition $M_0(t)=\sum_k N_k$ is just the total number of nodes in
the network.  It is clear by the nature of the growth process that this
quantity simply grows as $t$.  In a similar fashion, the first moment of the
degree distribution obeys $\dot M_1(t)=2$ with solution $M_1(t)= M_1(0)+2t$.
This time evolution for $M_1$ can be understood either by explicitly summing
the rate equations, or by observing that this first moment simply equals the
total number of link endpoints.  Clearly, this quantity must grow as $2t$
since the introduction of a single node introduces two link endpoints.  Thus
we find the simple result that the first two moments are {\em independent\/}
of the attachment kernel $A_k$ and grow {\em linearly} with time.  On the
other hand, higher moments and the degree distribution itself do depend in an
essential way on the kernel $A_k$.

As a preview to the general behavior for the degree distribution, consider
the strictly linear kernel \cite{BA,KRL,DMS}, for which $A(t)$ coincides with
$M_1(t)$.  In this case, we can solve Eqs.~(\ref{Nk}) for an arbitrary
initial condition.  However, since the long-time behavior is most
interesting, we limit ourselves to the asymptotic regime ($t\to\infty$) where
the initial condition is irrelevant.  Using therefore $M_1=2t$, we solve the
first few of Eqs.~(\ref{Nk}) directly and obtain $N_1=2t/3$, $N_2=t/6$, {\it
  etc}.  Thus each of the $N_k$ grow linearly with time.  Accordingly, we
substitute $N_k(t)=t\,n_k$ in Eqs.~(\ref{Nk}) to yield the simple recursion
relation $n_k=n_{k-1} (k-1)/(k+2)$.  Solving for $n_k$ gives
\begin{equation}
\label{nk1}
n_k={4\over k(k+1)(k+2)}. 
\end{equation}

Returning to the case of general attachment kernels, let us assume that the
degree distribution and $A(t)$ both grow linearly with time.  This hypothesis
can be easily verified numerically for attachment kernels that do not grow
faster than linearly with $k$.  Then substituting $N_k(t)=t\,n_k$ and
$A(t)=\mu t$ into Eqs.~(\ref{Nk}) we obtain the recursion relation
$n_k=n_{k-1} A_{k-1}/(\mu+A_k)$ and $n_1=\mu/(\mu+A_1)$.  Finally, solving
for $n_k$, we obtain the formal expression
\begin{equation}
\label{Nkgen}
n_k={\mu\over A_k}\prod_{j=1}^{k}
\left(1+{\mu\over A_j}\right)^{-1}.
\end{equation}
To complete the solution, we need the amplitude $\mu$.  Using the definition
$\mu=\sum_{j\geq 1}A_jn_j$ in Eq.~(\ref{Nkgen}), we obtain the implicit
relation
\begin{equation}
\label{mugen}
\sum_{k=1}^\infty \prod_{j=1}^{k}
\left(1+{\mu\over A_j}\right)^{-1}=1
\end{equation}
which shows that the amplitude $\mu$ depends on the entire attachment kernel.  

For the generic case $A_k\sim k^\gamma$, we substitute this form into
Eq.~(\ref{Nkgen}) and then rewrite the product as the exponential of a sum of
a logarithm.  In the continuum limit, we convert this sum to an integral,
expand the logarithm to lowest order, and then evaluate the integral to yield
the following basic results:
\begin{eqnarray}
\label{cases} 
n_k\sim\cases{
k^{-\gamma}\exp
\left[-\mu\left({{k^{1-\gamma}-2^{1-\gamma}}\over 1-\gamma}\right)\right],
&$0\leq\gamma<1$;\cr
k^{-\nu}, \quad \nu>2,
& $\gamma=1$;\cr
{\rm best\ seller} & $1<\gamma<2$;\cr
{\rm bible} & $2<\gamma$.}
\end{eqnarray}

Thus the degree distribution decays exponentially for $\gamma=0$, as in the
case of the random graph, while for all $0<\gamma<1$, the distribution
exhibits robust stretched exponential behavior.  The linear kernel is the
case that has garnered much of the current research interest.  As shown
above, $n_k={4/[k(k+1)(k+2)]}$ for the strictly linear kernel $A_k=k$.  One
might anticipate that $n_k\propto k^{-3}$ holds for all {\em asymptotically}
linear kernels, $A_k\sim k$.  However, the situation is more delicate and the
degree distribution exponent depends on microscopic details of $A_k$.  {}From
Eq.~(\ref{Nkgen}), we obtain $n_k\sim k^{-\nu}$, where the exponent
$\nu=1+\mu$ can be tuned to {\em any} value larger than 2 \cite{KRL,KR}.
This non-universal behavior shows that one must be cautious in drawing
general conclusions from the GN with a linear attachment kernel.  


\begin{figure}[ht]
  \begin{center}
    \includegraphics[width=0.25\textwidth]{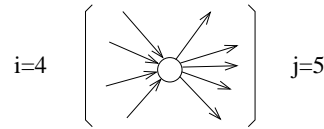}
 \caption{A node with in-degree $i=4$, out-degree $j=5$, and total
  degree 9.}~\label{degrees}
  \end{center}
\end{figure}

As an illustrative example of the vagaries of asymptotically linear kernels,
consider the shifted linear kernel $A_k=k+w$.  One way to motivate this
kernel is to explicitly keep track of link directionality.  In particular,
the node degree for an undirected graph naturally generalizes to the
in-degree and out-degree for a directed graph, the number of incoming and
outgoing links at a node, respectively.  Thus the total degree $k$ in a
directed graph is the sum of the in-degree $i$ and out-degree $j$
(Fig.~\ref{degrees}).  (More details on this model are given in the next
section.)~ The most general linear attachment kernel for a directed graph has
the form $A_{ij}=ai+bj$.  The GN corresponds to the case where the out-degree
of any node equals one; thus $j=1$ and $k=i+1$.  For this example the general
linear attachment kernel reduces to $A_k=a(k-1)+b$.  Since the overall scale
is irrelevant, we can re-write $A_k$ as the shifted linear kernel $A_k=k+w$,
with $w=-1+b/a$ that can vary over the range $-1<w<\infty$.

To determine the degree distribution for the shifted linear kernel, note that
$A(t)=\sum_jA_jN_j(t)$ simply equals \hbox{$A(t)=M_1(t)+wM_0(t)$}.  {}Using
$A=\mu t$, $M_0=t$ and $M_1=2t$, we get $\mu=2+w$ and hence the relation
$\nu=1+\mu$ from the previous paragraph becomes $\nu=3+w$.  Thus a simple
additive shift in the attachment kernel profoundly affects the asymptotic
degree distribution.  Furthermore, from Eq.~(\ref{Nkgen}) we determine the
entire degree distribution to be
\begin{equation}
\label{nkw}
n_k=(2+w)\,{\Gamma(3+2w)\over \Gamma(1+w)}\,
{\Gamma(k+w)\over \Gamma(k+3+2w)}.
\end{equation}

Finally, we outline the intriguing behavior for super-linear kernels.  In
this case, there is a ``runaway'' or gelation-like phenomenon in which one
node links to almost every other node.  For $\gamma>2$, all but a finite
number of nodes are linked to a {\em single} node that has the rest of the
links.  We term such an overwhelmingly popular node as a ``bible''.  For
$1<\gamma\leq 2$, the number of nodes with a just a few links is no longer
finite, but grows slower than linearly in time, and the remainder of the
nodes are linked to an extremely popular node that we now term ``best
seller''.  Full details about this runaway behavior are given in \cite{KRL}.

As a final parenthetical note, when the attachment kernel has the form
$A_k\propto k^\gamma$, with $\gamma<0$, there is preferential attachment to
poorly-connected sites.  Here, the degree distribution exhibits faster than
exponential decay, $n_k\propto k^{-\gamma(k-1)}$.  When $\gamma< -2$, the
propensity for avoiding popularity is so strong that there is a finite
probability of forming a ``worm'' graph in which each node attaches only to
its immediate predecessor.

\subsection{Degree Distribution of a Heterogeneous Network}

A practically-relevant generalization of the GN is to endow each node with an
intrinsic and permanently defined ``attractiveness'' \cite{BiA}.  This
accounts for the obvious fact that not all nodes are equivalent, but that
some are clearly more attractive than others at their inception.  Thus the
subsequent attachment rate to a node should be a function of both its degree
and its intrinsic attractiveness.  For this generalization, the rate equation
approach yields complete results with minimal additional effort beyond that
needed to solve the homogeneous network.

Let us assign each node an attractiveness parameter $\eta>0$, with arbitrary
distribution, at its inception.  This attractiveness modifies the node
attachment rate as follows: for a node with degree $k$ and attractiveness
$\eta$, the attachment rate is simply $A_k(\eta)$.  Now we need to
characterize nodes both by their degree and their attractiveness -- thus
$N_k(\eta)$ is the number of nodes with degree $k$ and attractiveness $\eta$.
This joint degree-attractiveness distribution obeys the rate equation,
\begin{equation}
\label{Nk-het}
{d N_k(\eta)\over dt}=
{A_{k-1}(\eta) N_{k-1}(\eta)-A_k(\eta) N_k(\eta)\over A}+p_0(\eta)\delta_{k1}.
\end{equation}
Here $p_0(\eta)$ is the probability that a newly-introduced node has
attractiveness $\eta$, and the normalization factor $A=\int d\eta
\sum_{k}A_k(\eta)N_k(\eta)$.

Following the same approach as that used to analyze Eq.~(\ref{Nk}), we
substitute $A=\mu t$ and $n_k(\eta)=tN_k(\eta)$ into Eq.~(\ref{Nk-het}) to
obtain the recursion relation
\begin{equation}
\label{Nkgen-het}
n_k(\eta)=p_0(\eta){\mu\over A_k(\eta)}\prod_{j=1}^{k}
\left(1+{\mu\over A_j(\eta)}\right)^{-1}.
\end{equation}

For concreteness, consider the linear attachment kernel $A_k(\eta)=\eta k$.
Then applying the same analysis as in the homogeneous network, we find
\begin{equation}
\label{nk-het}
n_k(\eta)= {\mu\,p_0(\eta)\over \eta}\,
{\Gamma(k)\, \Gamma\left(1+{\mu\over \eta}\right)\over  
\Gamma\left(k+1+{\mu\over \eta}\right)}.
\end{equation}
To determine the amplitude $\mu$ we substitute (\ref{nk-het}) into the
definition $\mu=\int d\eta\, \sum_{k\geq 1}A_k(\eta)\,n_k(\eta)$ and use the
identity \cite{knuth}
\begin{eqnarray*}
\label{identity}
\sum_{k=1}^\infty {\Gamma(k+u)\over \Gamma(k+v)}
={\Gamma(u+1)\over (v-u-1)\,\Gamma(v)}
\end{eqnarray*}
to simplify the sum.  This yields the implicit relation
\begin{equation}
\label{mu-het}
1=\int d\eta\, p_0(\eta)\,\left({\mu\over \eta}-1\right)^{-1}.
\end{equation}
This condition on $\mu$ leads to two alternatives: If the support of $\eta$
is unbounded, then the integral diverges and there is no solution for $\mu$.
In this limit, the most attractive node is connected to a finite fraction of
all links.  Conversely, if the support of $\eta$ is bounded, the resulting
degree distribution is similar to that of the homogeneous network.  For fixed
$\eta$, $n_k(\eta)\sim k^{-\nu(\eta)}$ with an attractiveness-dependent decay
exponent $\nu(\eta)=1+\mu/\eta$.  Amusingly, the total degree distribution
$n_k=\int d\eta\,n_k(\eta)$ is no longer a strict power law \cite{BiA}.
Rather, the asymptotic behavior is governed by properties of the initial
attractiveness distribution near the upper cutoff.  In particular, if
$p_0(\eta)\sim (\eta_{\rm max}-\eta)^{\omega-1}$ (with $\omega>0$ to ensure
normalization), the total degree distribution exhibits a logarithmic
correction
\begin{equation}
\label{nk-asymp-het}
n_k\sim k^{-(1+\mu/\eta_{\rm max})}\,(\ln k)^{-\omega}. 
\end{equation}

\subsection{Age Distribution}

In addition to the degree distribution, we determine {\em when} connections
occur.  Naively, we expect that older nodes will be better connected.  We
study this feature by resolving each node both by its degree and its age to
provide a more complete understanding of the network evolution.  Thus define
$c_k(t,a)$ to be the average number of nodes of age $a$ that have $k-1$
incoming links at time $t$.  Here age $a$ means that the node was introduced
at time $t-a$.  The original degree distribution may be recovered from the
joint age-degree distribution through $N_k(t)=\int_0^t da\,c_k(t,a)$.

For simplicity, we consider only the case of the strictly linear kernel; more
general kernels were considered in Ref.~\cite{KR}.  The joint age-degree
distribution evolves according to the rate equation
\begin{equation}
\label{ck1}
\left({\partial \over \partial t}+{\partial \over \partial a}\right)c_k 
={A_{k-1}c_{k-1}-A_k c_k\over 2t}
+\delta_{k1}\delta(a).
\end{equation}
The second term on the left accounts for the aging of nodes.  We assume here
that the probability of linking to a given node again depends only on its
degree and not on its age.  Finally, we again have used $A(t)\equiv
M_1(t)\simeq 2t$ for the linear attachment kernel in the long-time limit.

The homogeneous form of this equation implies that solution should be
self-similar.  Thus we seek a solution as a function of the {\em single}
variable $a/t$ rather than two separate variables.  Writing
$c_k(t,a)=f_k(x)$ with $x=1-{a\over t}$, we convert Eq.~(\ref{ck1}) into the
ordinary differential equation
\begin{equation}
\label{fk1}
-2x\,{df_k\over dx}=(k-1) f_{k-1}-k f_k.
\end{equation}
We omit the delta function term, since it merely provides the boundary
condition $c_k(t,a=0)=\delta_{k1}$, or $f_k(1)=\delta_{k1}$.

The solution to this boundary-value problem may be simplified by assuming the
exponential solution $f_k=\Phi\varphi^{k-1}$; this is consistent with the
boundary condition, provided that $\Phi(1)=1$ and $\varphi(1)=0$.  This
ansatz reduces the infinite set of rate equations (\ref{fk1}) into two
elementary differential equations for $\varphi(x)$ and $\Phi(x)$ whose
solutions are $\varphi(x)=1-\sqrt{x}$ and $\Phi(x)=\sqrt{x}$.  In terms of
the original variables of $a$ and $t$, the joint age-degree distribution is
then
\begin{eqnarray}
\label{ck1all}
c_k(t,a)=\sqrt{1-{a\over t}}\left\{1-\sqrt{1-{a\over t}}\right\}^{k-1}.
\end{eqnarray}

\begin{figure}[ht]
  \begin{center}
    \includegraphics[width=0.4\textwidth]{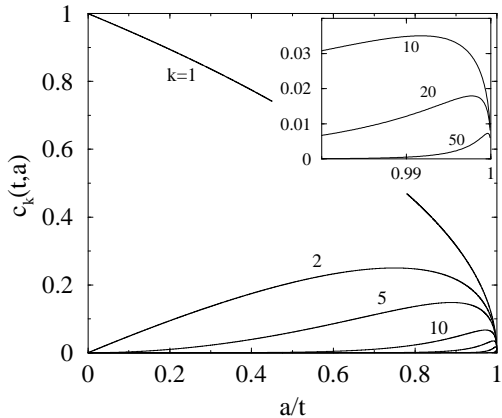} 
\caption{Age-dependent degree distribution for the GN for the linear 
  attachment kernel.  Low-degree nodes tend to be relatively young while
  high-degree nodes are old.  The inset shows detail for $a/t\geq 0.98$.}
\label{age}
\end{center}
\end{figure}

Thus the degree distribution for fixed-age nodes decays {\em exponentially},
with a characteristic degree that diverges as $\langle k\rangle\sim
(1-a/t)^{-1/2}$ for $a\to t$.  As expected, young nodes (those with $a/t\to
0$) typically have a small degree while old nodes have large degree
(Fig.~\ref{age}).  It is the large characteristic degree of old nodes that
ultimately leads to a {\em power-law} total degree distribution when the
joint age-degree distribution is integrated over all ages.

\subsection{Node Degree Correlations}

The rate equation approach is sufficiently versatile that we can also obtain
much deeper geometrical properties of growing networks.  One such property is
the correlation between degrees of connected nodes \cite{KR}.  These develop
naturally because a node with large degree is likely to be old.  Thus its
ancestor is also old and hence also has a large degree.  In the context of
the web, this correlation merely expresses that obvious fact that it is more
likely that popular web sites have hyperlinks among each other rather than to
marginal sites.

\begin{figure}[ht]
  \begin{center}
    \includegraphics[width=0.2\textwidth]{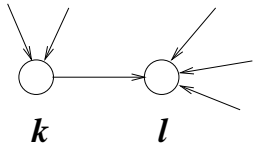}
\caption{Definition of the node degree correlation $C_{kl}$ for the case
  $k=3$ and $l=4$.}
\label{corr-def}
\end{center}
\end{figure}

To quantify the node degree correlation, we define $C_{kl}(t)$ as the number
of nodes of degree $k$ that attach to an ancestor node of degree $l$
(Fig.~\ref{corr-def}).  For example, in the network of Fig.~\ref{network},
there are $N_1=6$ nodes of degree 1, with $C_{12}=C_{13}=C_{15}=2$.  There
are also $N_2=2$ nodes of degree 2, with $C_{25}=2$, and $N_3=1$ nodes of
degree 3, with $C_{35}=1$.

\begin{figure}[ht]
  \begin{center}
    \includegraphics[width=0.7\textwidth]{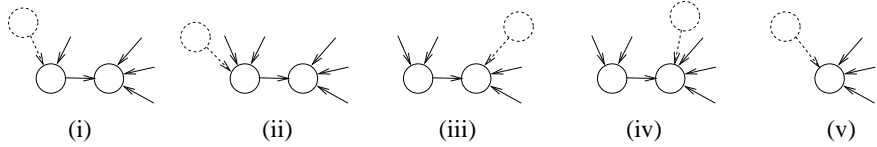}
\caption{The processes that contribute ((i)--(v) in order) 
  to the various terms in the rate equation (\ref{Nkl}).  The newly-added
  node and link are shown dashed.}
\label{corr-RE}
\end{center}
\end{figure}

For simplicity, we again specialize to the case of the strictly linear
attachment kernel.  More general kernels can also be treated within our
general framework \cite{KR}.  For the linear attachment kernel, the degree
correlation $C_{kl}(t)$ evolves according to the rate equation
\begin{eqnarray}
\label{Nkl}
M_1\,{d C_{kl}\over dt}=(k-1) C_{k-1,l}-kC_{kl}+
(l-1) C_{k,l-1}-l C_{kl}+(l-1)C_{l-1}\,\delta_{k1}.
\end{eqnarray}
The processes that gives rise to each term in this equation are illustrated in
Fig.~\ref{corr-RE}.  The first two terms on the right account for the change
in $C_{kl}$ due to the addition of a link onto a node of degree $k-1$ (gain)
or $k$ (loss) respectively, while the second set of terms gives the change in
$C_{kl}$ due to the addition of a link onto the ancestor node.  Finally, the
last term accounts for the gain in $C_{1l}$ due to the addition of a new
node.

As in the case of the node degree, the time dependence can be separated as
$C_{kl}= tc_{kl}$.  This reduces Eqs.~(\ref{Nkl}) to the time-independent
recursion relation,
\begin{eqnarray}
\label{nkl}
(k+l+2)c_{kl}=(k-1) c_{k-1,l}+(l-1) c_{k,l-1}
+(l-1)c_{l-1}\,\delta_{k1}.
\end{eqnarray}
This can be further reduced to a constant-coefficient inhomogeneous recursion
relation by the substitution
\begin{eqnarray*}
\label{Akl}
c_{kl}={\Gamma(k)\,\Gamma(l)\over \Gamma(k+l+3)}\,\,d_{kl}
\end{eqnarray*}
to yield
\begin{equation}
\label{A}
d_{kl}=d_{k-1,l}+d_{k,l-1}+4(l+2)\delta_{k1}.
\end{equation}
Solving Eqs.~(\ref{A}) for the first few $k$ yields the pattern of dependence
on $k$ and $l$ from which one can then infer the solution
\begin{equation}
\label{A-sol}
d_{kl}=4\,{\Gamma(k+l)\over \Gamma(k+2)\,\Gamma(l-1)}
+12\,{\Gamma(k+l-1)\over \Gamma(k+1)\,\Gamma(l-1)},
\end{equation}
from which we ultimately obtain
\begin{eqnarray}
\label{nkl-sol}
c_{kl}={4(l-1)\over k(k+l)(k+l+1)(k+l+2)}\left[{1\over k+1}
+{3\over k+l-1}\right].
\end{eqnarray}
The important feature of this result is that the joint distribution does not
factorize, that is, $c_{kl}\ne n_kn_{l}$.  This correlation between the
degrees of connected nodes is an important distinction between the GN and
classical random graphs.

While the solution of Eq.~(\ref{nkl-sol}) is unwieldy, it greatly simplifies
in the scaling regime, $k\to\infty$ and $l\to\infty$ with $y=l/k$ finite.
The scaled form of the solution is
\begin{eqnarray}
\label{nkl-scal}
c_{kl}=k^{-4}\,{4y(y+4)\over (1+y)^4}.
\end{eqnarray}
For fixed large $k$, the distribution $c_{kl}$ has a single maximum at
$y^*=(\sqrt{33}-5)/2 \cong 0.372$.  Thus a node whose degree $k$ is large is
typically linked to another node whose degree is also large; the typical
degree of the ancestor is 37\% that of the daughter node.  In general, when
$k$ and $l$ are both large and their ratio is different from one, the
limiting behaviors of $c_{kl}$ are
\begin{equation}
\label{nklext}
c_{kl}\to\cases{16\,(l/k^5)    & $l\ll k$,\cr
                4/(k^2\,l^2)   & $l\gg k$.\cr}
\end{equation}
Here we explicitly see the absence of factorization in the degree
correlation: $c_{kl}\ne n_kn_{l}\propto (k\,l)^{-3}$.

\subsection{Global Properties}

\begin{figure}[ht]
  \begin{center}
    \includegraphics[width=0.35\textwidth]{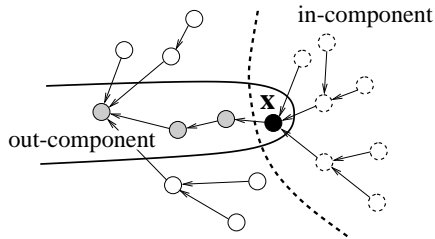}
 \caption{In-component and out-components of node {\bf x}.}~\label{in-out}
  \end{center}
\end{figure}

In addition to elucidating the degree distribution and degree correlations,
the rate equations can be applied to determine global properties.  One useful
example is the {\em out-component\/} with respect to a given node {\bf x} --
this is the set of nodes that can be reached by following directed links that
emanate from {\bf x} (Fig.~\ref{in-out}).  In the context of the web, this is
the set of nodes that are reached by following hyperlinks that emanate from a
fixed node to target nodes, and then iteratively following target nodes ad
infinitum.  In a similar vein, one may enumerate all nodes that refer to a
fixed node, plus all nodes that refer these daughter nodes, {\it etc}.  This
progeny comprises the in-component to node {\bf x} -- the set from which {\bf
  x} can be reached by following a path of directed links.

\subsubsection{The In-Component}

For simplicity, we study the in-component size distribution for the GN with a
constant attachment kernel, $A_k=1$.  We consider this kernel because many
results about network components are {\it independent\/} of the form of the
kernel and thus it suffices to consider the simplest situation; the extension
to more general attachment kernels is discussed in \cite{KR}.

For the constant attachment kernel, the number $I_s(t)$ of in-components with
$s$ nodes satisfies the rate equation
\begin{equation}
\label{Ik}
{d I_s\over dt}={(s-1)I_{s-1}-sI_s\over A}+\delta_{s1}.
\end{equation}
The loss term accounts for processes in which the attachment of a new node to
an in-component of size $s$ increases its size by one.  This gives a loss
rate that is proportional to $s$.  If there is more than one in-component of
size $s$ they must be disjoint, so that the total loss rate for $I_s(t)$ is
simply $sI_s(t)$.  A similar argument applies for the gain term.  Finally,
dividing by $A(t)=\sum_j A_j N_j(t)$ converts these rates to normalized
probabilities.  For the constant attachment kernel, $A(t)=M_0(t)$, so
asymptotically $A=t$.  Interestingly, Eq.~(\ref{Ik}) is almost identical to
the rate equations for the degree distribution for the GN with linear
attachment kernel, except that the prefactor equals $t^{-1}$ rather than
$(2t)^{-1}$.  This change in the normalization factor is responsible for
shifting the exponent of the resulting distribution from $-3$ to $-2$.

To determine $I_s(t)$, we again note, by explicitly solving the first few of
the rate equations, that each $I_s$ grows linearly in time.  Thus we
substitute $I_s(t)=ti_s$ into Eqs.~(\ref{Ik}) to obtain $i_1=1/2$ and
$i_s=i_{s-1}(s-1)/(s+1)$.  This immediately gives
\begin{equation}
\label{is}
i_s={1\over s(s+1)}.  
\end{equation}
This $s^{-2}$ tail for the in-component distribution is a robust feature,
{\em independent\/} of the form of the attachment kernel \cite{KR}.  This
$s^{-2}$ tail also agrees with recent measurements of the web \cite{www2}.

\subsubsection{The Out-Component}

\begin{figure}[ht]
  \begin{center}
    \includegraphics[width=0.35\textwidth]{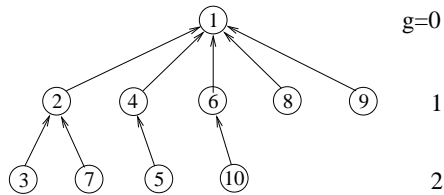}
 \caption{Genealogy of the network in Fig.~\ref{network}.  
   The nodes indices indicate when each is introduced.  The nodes are also
   arranged according to generation number.}~\label{genealogy}
  \end{center}
\end{figure}

The complementary out-component from each node can be determined by
constructing a mapping between the out-component and an underlying network
``genealogy''.  We build a genealogical tree for the GN by taking generation
$g=0$ to be the initial node.  Nodes that attach to those in generation $g$
form generation $g+1$; the node index does not matter in this
characterization.  For example, in the network of Fig.~\ref{network}, node 1
is the ``ancestor'' of 6, while 10 is the ``descendant'' of 6 and there are 5
nodes in generation $g=1$ and 4 in $g=2$.  This leads to the genealogical
tree of Fig.~\ref{genealogy}.

The genealogical tree provides a convenient way to characterize the
out-component distribution.  As one can directly verify from
Fig.~\ref{genealogy}, the number $O_s$ of out-components with $s$ nodes
equals $L_{s-1}$, the number of nodes in generation $s-1$ in the genealogical
tree.  We therefore compute $L_g(t)$, the size of generation $g$ at time $t$.
For this discussion, we again treat only the constant attachment kernel and
refer the reader to Ref.~\cite{KR} for more general attachment kernels.  We
determine $L_g(t)$ by noting that $L_g(t)$ increases when a new node attaches
to a node in generation $g-1$.  This occurs with rate $L_{g-1}/M_0$, where
$M_0(t)=1+t$ is the number of nodes.  This gives the differential equation
for $\dot L_g(t)=L_{g-1}/(1+t)$ with solution $L_g(\tau)={\tau^g/g!}$, where
$\tau=\ln(1+t)$.  Thus the number $O_s$ of out-components with $s$ nodes
equals
\begin{equation}
\label{Rk}
O_s(\tau)={\tau^{s-1}/ (s-1)!}.
\end{equation}
Note that the generation size $L_g(t)$ grows with $g$, when $g<\tau$, and
then decreases and becomes of order 1 when $g=e\tau$.  The genealogical tree
therefore contains approximately $e\tau$ generations at time $t$.  This
result allows us to determine the diameter of the network, since the maximum
distance between any pair of nodes is twice the distance from the root to the
last generation.  Therefore the diameter of the network scales as
$2e\tau\approx 2e\ln N$; this is the same dependence on $N$ as in the random
graph \cite{bol,jan}.  More importantly, this result shows that the diameter
of the GN is always small -- ranging from the order of $\ln N$ for a constant
attachment kernel, to the order of one for super-linear attachment kernels.

\section{The Web Graph}

In the world-wide web, link directionality is clearly relevant, as hyperlinks
go {\em from} an issuing website {\em to} a target website but not vice
versa.  Thus to characterize the local graph structure more fully, the node
degree should be resolved into the {\em in-degree} -- the number of incoming
links to a node, and the complementary {\em out-degree} (Fig.~\ref{degrees}).
Measurements on the web indicate that these distributions are power laws with
different exponents \cite{www3}.  These properties can be accounted for by
the web graph (WG) model (Fig.~\ref{io-growth}) and the rate equations
provide an extremely convenient analysis tool.

\subsection{Average Degrees}

Let us first determine the average node degrees (in-degree, out-degree, and
total degree) of the WG.  Let $N(t)$ be the total number of nodes, and $I(t)$
and $J(t)$ the in-degree and out-degree of the entire network, respectively.
According to the elemental growth steps of the model, these degrees evolve by
one of the following two possibilities:
\begin{eqnarray*}
(N,I,J)\to \cases{(N+1,I+1,J+1)  & with probability $p$,\cr
                  (N,I+1,J+1)    & with probability $q$.}  
\end{eqnarray*}
That is, with probability $p$ a new node and new directed link are created
(Fig.~\ref{io-growth}) so that the number of nodes and both the total in- and
out-degrees increase by one.  Conversely, with probability $q$ a new directed
link is created and the in- and out-degrees each increase by one, while the
total number of nodes is unchanged.  As a result, $N(t)=pt$, and
$I(t)=J(t)=t$.  Thus the average in- and out-degrees, ${\cal D}_{\rm in}\equiv
I(t)/N(t)$ and ${\cal D}_{\rm out}\equiv J(t)/N(t)$, are both equal to $1/p$.

\subsection{Degree Distributions}

To determine the degree distributions, we need to specify: (i) the {\em
  attachment rate} $A(i,j)$, defined as the probability that a
newly-introduced node links to an existing node with $i$ incoming and $j$
outgoing links, and (ii) the {\em creation rate} $C(i_1,j_1|i_2,j_2)$,
defined as the probability of adding a new link from a $(i_1,j_1)$ node to a
$(i_2,j_2)$ node.  We will use rates that are expected to occur in
the web.  Clearly, the attachment and creation rates should be non-decreasing
in $i$ and $j$.  Moreover, it seems intuitively plausible that the attachment
rate depends only on the in-degree of the target node, $A(i,j)=A_i$; {\it
  i.e.}, a website designer decides to create link to a target based only on
the popularity of the latter.  In the same spirit, we take the link creation
rate to depend only on the out-degree of the issuing node and the in-degree
of the target node, $C(i_1,j_1|i_2,j_2)= C(j_1,i_2)$.  The former property
reflects the fact that the development rate of a site depends only on the
number of outgoing links.

The interesting situation of power-law degree distributions arises for
asymptotically linear rates, and we therefore consider
\begin{equation}
\label{AC}
A_i=i+\lambda_{\rm in} \qquad{\rm and}\qquad C(j,i)=(i+\lambda_{\rm
  in})(j+\lambda_{\rm out})
\end{equation}
The parameters $\lambda_{\rm in}$ and $\lambda_{\rm out}$ must satisfy the
constraint $\lambda_{\rm in}>0$ and $\lambda_{\rm out}>-1$ to ensure that the
rates are positive for all attainable in- and out-degree values, $i\geq 0$
and $j\geq 1$.

With these rates, the joint degree distribution, $N_{ij}(t)$, defined as the
average number of nodes with $i$ incoming and $j$ outgoing links, evolves
according to
\begin{eqnarray}
\label{Nij}
{d N_{ij}\over dt}&=&
(p+q)\left[{(i-1+\lambda_{\rm in})N_{i-1,j}
-(i+\lambda_{\rm in})N_{ij}\over I+\lambda_{\rm in} N}\right]\\
 &&\hskip 0.285in 
+q\left[{(j-1+\lambda_{\rm out})N_{i,j-1}
-(j+\lambda_{\rm out})N_{ij}\over J+\lambda_{\rm out} N}\right]
+p\,\delta_{i0}\delta_{j1}.\nonumber
\end{eqnarray}
The first group of terms on the right accounts for the changes in the
in-degree of target nodes by simultaneous creation of a new node and link
(probability $p$) or by creation of a new link only (probability $q$).  For
example, the creation of a link to a node with in-degree $i$ leads to a loss
in the number of such nodes.  This occurs with rate $(p+q)(i+\lambda_{\rm
  in})N_{ij}$, divided by the appropriate normalization factor
$\sum_{i,j}(i+\lambda_{\rm in})N_{ij}= I+\lambda_{\rm in} N$.  The factor
$p+q=1$ in Eq.~(\ref{Nij}) is explicitly written to make clear these two
types of processes.  Similarly, the second group of terms account for
out-degree changes.  These occur due to the creation of new links between
already existing nodes -- hence the prefactor $q$.  The last term accounts
for the introduction of new nodes with no incoming links and one outgoing
link.  As a useful consistency check, one may verify that the total number of
nodes, $N=\sum_{i,j} N_{ij}$, grows according to $\dot N=p$, while the total
in- and out-degrees, $I=\sum_{i,j} iN_{ij}$ and $J=\sum_{i,j} jN_{ij}$, obey
$\dot I=\dot J=1$.

By solving the first few of Eqs.~(\ref{Nij}), it is again clear that the
$N_{ij}$ grow linearly with time.  Accordingly, we substitute
$N_{ij}(t)=t\,n_{ij}$, as well as $N=pt$ and $I=J=t$, into Eqs.~(\ref{Nij})
to yield a recursion relation for $n_{ij}$.  Using the shorthand notations,
\begin{eqnarray*}
a=q\,{1+p\lambda_{\rm in}\over 1+p\lambda_{\rm out}}\quad {\rm and}\quad 
b=1+(1+p)\lambda_{\rm in},
\end{eqnarray*}
the recursion relation for $n_{ij}$ is
\begin{eqnarray}
\label{nij}
[i+a(j+\lambda_{\rm out})+b]n_{ij}
=(i-1+\lambda_{\rm in})n_{i-1,j}+a(j-1+\lambda_{\rm out})n_{i,j-1}
+p(1+p\lambda_{\rm in})\delta_{i0}\delta_{j1}.
\end{eqnarray}
The in-degree and out-degree distributions are straightforwardly expressed
through the joint distribution: ${\cal I}_i(t)
=\sum_j N_{ij}(t)$ and ${\cal O}_j(t)=\sum_i N_{ij}(t)$.  Because of the
linear time dependence of the node degrees, we write ${\cal I}_i(t)=t\,I_i$
and ${\cal O}_j(t)=t\,O_j$.  The densities $I_i$ and $O_j$ satisfy
\begin{subeqnarray}
\label{Ii}
(i+b)I_{i} &=&(i-1+\lambda_{\rm in})I_{i-1}
+p(1+p\lambda_{\rm in})\delta_{i0},\\
\left(j+{1\over q}+{\lambda_{\rm out}\over q}\right)O_j 
&=&(j-1+\lambda_{\rm out})O_{j-1}+p{1+p\lambda_{\rm out}\over q}\delta_{j1},
\end{subeqnarray}
respectively.  The solution to these recursion formulae may be expressed
in terms of the following ratios of gamma functions
\begin{subeqnarray}
\label{I-sol}
I_{i}&=&I_0\,{\Gamma(i+\lambda_{\rm in})\,\Gamma(b+1)\over
\Gamma(i+b+1)\,\Gamma(\lambda_{\rm in})},\\
\label{O-sol}
O_{j}&=&O_1\,{\Gamma(j+\lambda_{\rm out})\,\,
\Gamma(2+q^{-1}+\lambda_{\rm out} q^{-1})\over 
\Gamma(j+1+q^{-1}+\lambda_{\rm out} q^{-1})\,\Gamma(1+\lambda_{\rm out})},
\end{subeqnarray}
with $I_0=p(1+p\lambda_{\rm in})/b$ and
$O_1=p(1+p\lambda_{\rm out})/(1+q+\lambda_{\rm out})$.

{}From the asymptotics of the gamma function, the asymptotic behavior of the
in- and out-degree distributions have the distinct power law forms \cite{KRR},
\begin{subeqnarray}
\label{in}
I_i\sim i^{-\nu_{\rm in}},~~~\qquad \nu_{\rm in}&=&2+p\lambda_{\rm in},\\
\hskip 0.7in O_j\sim j^{-\nu_{\rm out}}, \qquad
\nu_{\rm out}&=&1+q^{-1}+\lambda_{\rm out}\, pq^{-1},
\end{subeqnarray}
with $\nu_{\rm in}$ and $\nu_{\rm out}$ both necessarily greater than 2.  Let
us now compare these predictions with current data for the web \cite{www3}.
First, the value of $p$ is fixed by noting that $p^{-1}$ equals the average
degree of the entire network.  Current data for the web gives ${\cal D}_{\rm
  in}\equiv {\cal D}_{\rm out}\approx 7.5$, and thus we set $p^{-1}=0.75$.
Now Eqs.~(\ref{in}) contain two free parameters and by choosing them to be
$\lambda_{\rm in}=0.75$ and $\lambda_{\rm out}=3.55$ we reproduced the
observed exponents for the degree distributions of the web, $\nu_{\rm
  in}\approx 2.1$ and $\nu_{\rm out}\approx 2.7$, respectively.  The fact
that the parameters $\lambda_{\rm in}$ and $\lambda_{\rm out}$ are of the
order of one indicates that the model with linear rates of node attachment
and bilinear rates of link creation is a viable description of the web.

\section{Multicomponent Graph}

In addition to the degree distributions, current measurements indicate that
the web consists of a ``giant'' component that contains approximately 91\% of
all nodes, and a large number of finite components \cite{www3}.  The models
discussed thus far are unsuited to describe the number and size distribution
of these components, since the growth rules necessarily produce only a single
connected component.  In this section, we outline a simple modification of
the WG, the multicomponent graph (MG), that naturally produces many
components.  In this example, the rate equations now provide a comprehensive
characterization for the size distribution of the components.

In the MG model, we simply separate node and link creation steps.  Namely,
when a node is introduced it does not immediately attach to an earlier node,
but rather, a new node begins its existence as isolated and joins the network
only when a link creation event reaches the new node.  For the average
network degrees, this small modification already has a significant effect.
The number of nodes and the total in- and out-degrees of the network, $N,I,J$
now increase with time as $N=pt$ and $I=J=qt$.  Thus the in- and out-degrees
of each node are time independent and equal to $qp^{-1}$, while the total
degree is ${\cal D}=2q/p$.

As in the case of the WG model, we study the case of a bilinear link creation
rate given in Eq.~(\ref{AC}), with now $\lambda_{\rm in},\lambda_{\rm out}>0$
to ensure that $C(j,i)>0$ for all permissible in- and out-degrees, $i\geq 0$
and $j\geq 0$.

\subsection{Local Properties}

We study local characteristics by employing the same approach as in the WG
model.  We find that results differ only in minute details, {\it e.g.}, the
in- and out-degree densities $I_i$ and $O_j$ are again the ratios of gamma
functions, and the respective exponents are
\begin{equation}
\label{inout}
\nu_{\rm in}=2\left(1+{\lambda_{\rm in}\over {\cal D}}\right),\qquad  
\nu_{\rm out}=2\left(1+{\lambda_{\rm out}\over {\cal D}}\right). 
\end{equation}
Notice the decoupling -- the in-degree exponent is independent of
$\lambda_{\rm out}$, while $\nu_{\rm out}$ is independent of $\lambda_{\rm
  in}$. The expressions (\ref{inout}) are neater than their WG counterparts,
reflecting the fact that the governing rules of the MG model are more
symmetric.

To complement our discussion, we now outline the asymptotic behavior of the
joint in- and out-degree distribution.  Although this distribution defies
general analysis, we can obtain partial and useful information by fixing one
index and letting the other index vary.  An elementary but cumbersome
analysis yields following limiting behaviors
\begin{equation}
\label{extreme}
n_{ij}\sim\cases{i^{-\xi_{\rm in}}, & $1\ll i$;\cr
                j^{-\xi_{\rm out}}, & $1\ll j$;}
\end{equation}
with
\begin{eqnarray*}
\xi_{\rm in} &=&\nu_{\rm in}+{{\cal D}\over 2}\,
{(\nu_{\rm in}-1)(\nu_{\rm out}-2)\over \nu_{\rm out}-1}\\
\xi_{\rm out} &=&\nu_{\rm out}+{{\cal D}\over 2}\,
{(\nu_{\rm out}-1)(\nu_{\rm in}-2)\over \nu_{\rm in}-1}.
\end{eqnarray*}
 
We also can determine the joint degree distribution analytically in the
subset of the parameter space where $\nu_{\rm in}=\nu_{\rm out}$, {\it i.e.},
$\lambda_{\rm in}=\lambda_{\rm out}$.  In what follows, we therefore denote
$\lambda_{\rm in}=\lambda_{\rm out}\equiv \lambda$.  The resulting recursion
equation for the joint degree distribution is
\begin{eqnarray}
\label{nij*}
(i+j+1+\lambda+\lambda q^{-1})n_{ij}=(i-1+\lambda)n_{i-1,j}
+(j-1+\lambda)n_{i,j-1}+c\,\delta_{i,0}\,\delta_{j,0},
\end{eqnarray}
with $c=p(1+2\lambda/{\cal D})$.  Because the degrees $i$ and $j$ appear in
Eq.~(\ref{nij*}) with equal prefactors, the substitution
\begin{eqnarray*}
\label{mij}
n_{ij}={\Gamma(i+\lambda)\,\Gamma(j+\lambda)\over 
\Gamma(i+j+2+\lambda+\lambda q^{-1})}\,\,m_{ij}
\end{eqnarray*}
reduces Eqs.~(\ref{nij*}) into the constant-coefficient recursion relation 
\begin{equation}
\label{m}
m_{ij}=m_{i-1,j}+m_{i,j-1}+\mu\,\delta_{i,0}\,\delta_{j,1},  \qquad
{\rm with}\quad \mu=c\,{\Gamma(1+\lambda+\lambda q^{-1})\over 
\Gamma^2(\lambda)}.
\end{equation}
We solve Eq.~(\ref{m}) by employing the generating function technique.
Multiplying Eq.~(\ref{m}) by $x^iy^j$ and summing over all $i,j\geq 0$, we
find that the generating function ${\mathcal M}(x,y)=\sum_{i,j\geq
  0}m_{ij}x^iy^j$ equals $\mu/(1-x-y)$.  Expanding ${\mathcal M}(x,y)$ in $x$
yields $\mu \sum x^i/(1-y)^{i+1}$ which we then expand in $y$ by employing
the identity $(1-y)^{-i-1}=\sum_{j\geq 0} {i+j\choose i}y^{j}$. Finally, we
arrive at 
\begin{equation}
\label{mij-sol}
m_{ij}=\mu\,\,{\Gamma(i+j+1)\over \Gamma(i+1)\,\Gamma(j+1)},
\end{equation}
from which the joint degree distribution is
\begin{equation}
\label{nij-sol}
n_{ij}={\mu\,\Gamma(i+\lambda)\,\Gamma(j+\lambda)\,\Gamma(i+j+1)\over 
\Gamma(i+1)\,\Gamma(j+1)\,\Gamma(i+j+2+\lambda+\lambda q^{-1})}
\longrightarrow \mu\,
{(ij)^{\lambda-1}\over (i+j)^{1+\lambda+\lambda/q}}, 
\quad{\rm as}\quad i,j\to\infty. 
\end{equation}
Thus again, the in- and out-degrees of a node are correlated: $n_{ij}\ne
I_iO_j\sim i^{-\nu}j^{-\nu}$.

\subsection{Global Properties}

Let us now turn now to the distribution of connected components (clusters,
for brevity).  For simplicity, we consider models with undirected links.  Let
us first estimate the total number of clusters ${\cal N}$.  At each time
step, ${\cal N}\to {\cal N}+1$ with probability $p$, or ${\cal N}\to {\cal
  N}-1$ with probability $q$.  This implies
\begin{equation}
\label{N}
{\cal N}=(p-q)t.  
\end{equation}
The gain rate of ${\cal N}$ is exactly equal to $p$, while in the loss term
we ignore self-connections and tacitly assume that links are always created
between different clusters.  In the long-time limit, self-connections should
be asymptotically negligible when the total number of clusters grows with
time and no macroscopic clusters ({\it i.e.}, components that contain a
finite fraction of all nodes) arise.  

This assumption of no self-connections greatly simplifies the description of
the cluster merging process.  Consider two clusters (labeled by $\alpha=1,2$)
with total in-degrees $i_\alpha$, out-degrees $j_\alpha$, and number of nodes
$k_\alpha$.  When these clusters merge, the combined cluster is characterized
by
\begin{eqnarray*}
\label{12}
i=i_1+i_2+1,\qquad
j=j_1+j_2+1,\qquad
k=k_1+k_2.
\end{eqnarray*}
Thus starting with single-node clusters with $(i,j,k)=(0,0,1)$, the above
merging rule leads to clusters that always satisfy the constraint $i=j=k-1$.
Thus the size $k$ characterizes both the in-degree and out-degree of
clusters.

To simplify formulae without sacrificing generality, we consider the link
creation rate of Eq.~(\ref{AC}), with $\lambda_{\rm in}=\lambda_{\rm out}=1$.
Then the merging rate $W(k_1,k_2)$ of the two clusters is proportional to
$(i_1+k_1)(j_2+k_2)+(i_2+k_2)(j_1+k_1)$, or
\begin{eqnarray*}
\label{w}
W(k_1,k_2)=(2k_1-1)(2k_2-1).  
\end{eqnarray*}
Let $C(k,t)$ denotes the number of clusters of mass $k$.  This distribution
evolves according to
\begin{eqnarray}
\label{comp}
{dC(k,t)\over dt}=
{q\over t^2}\sum_{k_1+k_2=k} (2k_1-1)(2k_2-1)\,C(k_1,t)C(k_2,t)
-{2q\over t}\,(2k-1)C(k,t)+p\,\delta_{k,1},
\end{eqnarray}
The first set of terms account for the gain in $C(k,t)$ due to the
coalescence of clusters of size $k_1$ and $k_2$, with $k_1+k_2=k$.
Similarly, the second set of terms accounts for the loss in $C(k,t)$ due to
the coalescence of a cluster of size $k$ with any other cluster.  The last
term accounts for the input of unit-size clusters.  These rate equations are
similar to those of irreversible aggregation with product kernel \cite{agg}.
The primary difference is that we explicitly treat the number of clusters as
finite.

One can verify that the total number of nodes $N(t)=\sum k\,C(k,t)$ grows
with rate $p$ and that the total number of clusters ${\cal N}(t)=\sum C(k,t)$
grows with rate $p-q$, in agreement with Eq.~(\ref{N}).  Solving the first
few Eqs.~(\ref{comp}) shows again that $C(k,t)$ grow linearly with time.
Accordingly, we substitute $C(k,t)=t\,c_k$ into Eqs.~(\ref{comp}) to yield
the time-independent recursion relation
\begin{eqnarray}
\label{mk}
c_k=q\sum_{k_1+k_2=k} (2k_1-1)(2k_2-1)\,c_{k_1}c_{k_2}
-2q(2k-1)c_k+p\,\delta_{k,1}.
\end{eqnarray}

A giant component, {\it i.e.}, a cluster that contains a finite fraction of
all the nodes, emerges when the link creation rate exceeds a threshold value.
To determine this threshold, we study the moments of the cluster size
distribution ${\cal M}_n=\sum_{k\geq 1} k^n\,c_k$.  We already know that the
first two moments are ${\cal M}_0=p-q$ and ${\cal M}_1=p$.  We can obtain an
equation for the second moment by multiplying Eq.~(\ref{mk}) by $k^2$ and
summing over $k\geq 1$ to give ${\cal M}_2 =2q(2{\cal M}_2-{\cal M}_1)^2+p$.
When this equation has a real solution, ${\cal M}_2$ is finite.  The
solution is
\begin{equation}
\label{M2}
{\cal M}_2={1+8pq-\sqrt{1-16pq}\over 16 q}
\end{equation}
and gives, when $1-16pq=0$, to a threshold value $p_c=(2+\sqrt{3})/4$.  For
$1-16pq\geq 0$ ($p>p_c$) all clusters have finite size and the second moment
is finite.

In this steady-state regime, we can obtain the cluster size distribution by
introducing the generating function ${\cal C}(z)=\sum_{k=1}^\infty c_k z^k$
to convert Eq.~(\ref{mk}) into the differential equation
\begin{equation}
\label{Cz}
2z{\cal C}'(z)-{\cal C}(z)=1-\sqrt{1-[pz-{\cal C}(z)]/q}.
\end{equation}
The asymptotic behavior of the cluster size distribution can now be read off
from the behavior of the generating function in the $z\to 1$ limit.  In
particular, the power-law behavior
\begin{equation}
\label{asym}
c_k\sim {B\over k^\tau}\quad{\rm as} \quad k\to\infty 
\end{equation}
implies that the corresponding generating function has the form
\begin{equation}
\label{gen}
{\cal C}(z)={\cal M}_0+{\cal M}_1(z-1)
+{{\cal M}_2-{\cal M}_1\over 2}\,(z-1)^2+
B\Gamma(1-\tau)(1-z)^{\tau-1}+\ldots.
\end{equation}
Here the asymptotic behavior is controlled by the dominant singular term
$(1-z)^{\tau-1}$.  However, there are also subdominant singular terms and
regular terms in the generating function.  In Eq.~(\ref{gen}) we explicitly
included the three regular terms which ensure that the first three moments of
the cluster-size distribution are correctly reproduced, namely, ${\cal
  C}(1)={\cal M}_0$, ${\cal C}'(1)={\cal M}_1$, and ${\cal C}''(1)={\cal
  M}_2-{\cal M}_1$.

Finally, substituting Eq.~(\ref{gen}) into Eq.~(\ref{Cz}) we find that the
dominant singular terms are of the order of $(1-z)^{\tau-2}$.  Balancing all
contributions of this order in the equation determines the exponent of the
cluster size distribution to be
\begin{equation}
\label{tau}
\tau=1+{2\over 1-\sqrt{1-16pq}}.
\end{equation}
This exponent satisfies the bound $\tau>3$ and thus justifies using the
behavior of the second moment of the size distribution as the criterion to
find the threshold value $p_c$.

For $p\geq p_c$ there is no giant cluster and the cluster size distribution
has a power-law tail with $\tau$ given by Eq.~(\ref{tau}).  Intriguingly, the
power-law form holds for any value $p>p_c$.  This is in stark contrast to
all other percolation-type phenomena, where away from the threshold, there is
an exponential tail in cluster size distributions \cite{percolation}.  Thus
in contrast to ordinary critical phenomena, the entire range $p>p_c$ is
critical. 

As a corollary to the power-law tail of the cluster size distribution for
$p>p_c$, we can estimate the size of the largest cluster $k_{\rm max}$ to see
how ``finite'' it really is.  Using the extreme statistics criterion
$\sum_{k\geq k_{\rm max}}N\,c_k=1$ we obtain $k_{\rm max}\sim N^{1/(\tau-1)}$,
or
\begin{equation}
\label{kmax}
k_{\rm max}\sim N^{(1-\sqrt{1-16pq})/2}.
\end{equation}
This is very different from the corresponding behavior on the random graph,
where below the percolation threshold the largest component scales
logarithmically with the number of nodes.  Thus for the random graph, the
dependence of $k_{\rm max}(N)$ changes from $\ln N$ just below, to $N$, just
above the percolation threshold; for the MG, the change is much more gentle:
from $N^{1/2}$ to $N$.

These considerations suggest that the phase transition in the MG is
dramatically different from the percolation transition.  Very recently,
simplified versions of the MG were studied
\cite{clusters,kk,sam,kkkr,french}.  Numerical \cite{clusters} and analytical
\cite{sam,kkkr,french} evidence suggest that the size of the giant component
$G(p)$ near the threshold scales as
\begin{equation}
\label{giant}
G(p)\propto \exp\left(-\,{{\rm const.}\over\sqrt{p_c-p}}\right).
\end{equation}
Therefore, the phase transition of this dynamically grown network is of
infinite order since all derivatives of $G(p)$ vanish as $p\to p_c$.  In
contrast, static random graphs with any desired degree distribution
\cite{reed} exhibit a standard percolation transition
\cite{clusters,reed,chung,dani}.

\section{Summary}

In this paper, we have presented a statistical physics viewpoint on growing
network problems.  This perspective is strongly influenced by the phenomenon
of aggregation kinetics, where the rate equation approach has proved
extremely useful.  From the wide range of results that we were able to obtain
for evolving networks, we hope that the reader appreciates both the
simplicity and the power of the rate equation method for characterizing
evolving networks.  We quantified the degree distribution of the growing
network model and found a diverse range of phenomenology that depends on the
form of the attachment kernel.  At the qualitative level, a stretched
exponential form for the degree distribution should be regarded as
``generic'', since it occurs for an attachment kernel that is sub-linear in
node degree ({\it e.g.}, $A_k\sim k^\gamma$ with $\gamma<1$).  On the other
hand, a power-law degree distribution arises only for linear attachment
kernels, $A_k\sim k$.  However, this result is ``non-generic'' as the degree
distribution exponent now depends on the detailed form of the attachment
kernel.

We investigated extensions of the basic growing network to incorporate
processes that naturally occur in the development in the web.  In particular,
by allowing for link directionality, the full degree distribution naturally
resolves into independent in-degree and out-degree distributions.  When the
rates at which links are created are linear functions of the in- and
out-degrees of the terminal nodes of the link, the in- and out-degree
distributions are power laws with different exponents, $\nu_{\rm in}$ and
$\nu_{\rm out}$, that match with current measurements on the web with
reasonable values for the model parameters.  We also considered a model with
independent node and link creation rates.  This leads to a network with many
independent components and now the size distribution of these components is
an important characteristic.  We have characterized basic aspects of this
process by the rate equation approach and showed that the network is in a
critical state even away from the percolation threshold.  The rate equation
approach also provides evidence of an unusual, infinite-order percolation
transition.

While statistical physics tools have fueled much progress in elucidating the
structure of growing networks, there are still many open questions.  One set
is associated with understanding dynamical processes in such networks.  For
example, what is the nature of information transmission?  What governs the
formation of traffic jams on the web?  Another set is concerned with growth
mechanisms.  While we can make much progress in characterizing networks with
idealized growth rules, it is important to understand the actual rules that
govern the growth of the Internet.  These issues appear to be fruitful
challenges for future research.

\section{Acknowledgements} 

It is a pleasure to thank Francois Leyvraz and Geoff Rodgers for
collaborations that led to some of the work reported here.  We also thank
John Byers and Mark Crovella for numerous informative discussions.  Finally,
we are grateful to NSF grants INT9600232 and DMR9978902 for financial
support.

\end{document}